# Spectral calibration for deriving surface mineralogy of Asteroid (25143) Itokawa from Hayabusa Near-Infrared Spectrometer (NIRS) Data

Megha Bhatt
Planetary Science Institute, 1700 East Fort Lowell Road, Tucson, AZ 85719, USA
Email: mubhatt19@gmail.com

Vishnu Reddy
Planetary Science Institute, 1700 East Fort Lowell Road, Tucson, AZ 85719, USA

Lucille Le Corre
Planetary Science Institute, 1700 East Fort Lowell Road, Tucson, AZ 85719, USA

Juan A. Sanchez
Planetary Science Institute, 1700 East Fort Lowell, Suite 106, Tucson, AZ 85719, USA

Tasha Dunn
Department of Geology, Colby College, 4000 Mayflower Hill Dr, Waterville, ME 04901, USA

Matthew R. M. Izawa
University of Winnipeg, 515 Portage Ave, Winnipeg, MB R3B 2E9, Canada

Jian-Yang Li
Planetary Science Institute, 1700 East Fort Lowell Road, Tucson, AZ 85719, USA

Kris J. Becker
USGS Astrogeology Science Center, 2255 N Gemini Dr, Flagstaff, AZ 86001, USA

Lynn Weller
USGS Astrogeology Science Center, 2255 N Gemini Dr, Flagstaff, AZ 86001, USA

1700 East Fort Lowell Road, Suite 106
Tucson 85719
(808) 342-8932 (voice)
reddy@psi.edu

**Abstract**

We present spectral calibration equations for determining mafic silicate composition of near-Earth asteroid (25143) Itokawa from visible/near-infrared (VNIR) spectra (0.85-2.1 µm) measured using the Near Infrared Spectrometer (NIRS), on board the Japanese Hayabusa spacecraft. Itokawa was the target of the Hayabusa sample return mission and has a surface composition similar to LL-type ordinary chondrites. Exisitng laboratory spectral calibrations (e.g., Dunn et al. 2010) use a spectral wavelength range that is wider (0.75-2.5 µm) than that of the NIRS instrument (0.85-2.1 µm) making them unfit for interpreting the Hayabusa spectral data currently archived in the Planetary Data System (PDS). We used laboratory measured near-infrared (NIR) reflectance spectra of ordinary (H, L and LL) chondrites from the study of Dunn et al. (2010), which we resampled to the NIRS wavelength range. Using spectral parameters extracted from these resampled spectra we established a relationship between band parameters and their mafic silicate composition (olivine and low-Ca pyroxene). We found a correlation >90% between mafic silicate composition (fayalite and forsterite mol. %) estimated by our spectral method and X-ray diffraction (XRD) measured values. The standard deviation between the measured and estimated values is 1.5 and 1.1 mol. % for fayalite and ferrosilite, respectively. To test the validity of the newly derived equations we blind tested them using nine laboratory measured spectra of L and LL type chondrites with known composition. We found that the absolute difference between the measured and computed values is in the range 0.1 to 1.6 mol. %. Our study suggests that the derived calibration is robust and can be applied to Hayabusa NIRS data despite its limited spectral range (0.85-2.1 µm).  We applied the derived equations to a subset of uncalibrated NIRS spectra and the derived fayalite and ferrosilite values are consistent with Itokawa having a LL chondrite type surface composition. We intend to develop a photometric model to calibrate the NIRS spectra and apply the derived equations to create a global mineralogical map of Itokawa.

# 1. Introduction

The Hayabusa spacecraft was launched on 9 May 2003 for a rendezvous with Apollo-type near-Earth asteroid (NEA), (25143) Itokawa, for sample return to Earth. Itokawa's orthogonal axes are 535, 294, and 209 meters (Fujiwara et al., 2006) and a spectral signature consistent with an LL chondrite (Binzel et al. 2001). The spacecraft arrived at the asteroid on 12 September 2005 and spent three months collecting science data in a station-keeping, heliocentric orbit (Saito et al., 2006; Fujiwara et al., 2006). In November 2005, Hayabusa spacecraft performed Itokawa rendezvous operations including two touchdowns for the first asteroid sample collection (Tsuchiyama 2014). It successfully returned regolith dust particles of 10 to 300 microns sized grains from Itokawa to Earth in June 2010 (Tsuchiyama 2014). Analysis of the returned samples showed that Itokawa is made of re-accreted fragments of interior portions of a once larger asteroid of ~20 km in diameter (e.g., Nakamura et al. 2011, 2014); it thus represents only a small fraction of the original asteroid's mass. Currently, the Hayabusa mission is the only robotic mission that has successfully orbited, made multiple touchdowns and ascents from the surface of an asteroid, landed, and returned samples of a small NEA.

The Hayabusa instrument package included the Asteroid Multiband Imaging Camera (AMICA), Near Infrared Spectrometer (NIRS), X-ray Spectrometer (XRS), LIDAR altimeter (LIDAR), surface hopper (MINERVA) and sample collection and return systems (Saito et al.,2006; Abe et al. 2006; Okada et al., 2006; Mukai et al., 2006; Yoshimitsu et al., 2004; Nakamura et al. 2011). The multispectral AMICA and hyperspectral NIRS data were obtained by Hayabusa to help constrain composition and mineralogy of Itokawa's surface. The NIRS is a 64-channel grating type point spectrometer. This instrument operated in the wavelength range 0.76-2.25 μm in 64 spectral bands, with a spectral resolution of 24 nm, at a spatial resolution of approximately 17 x 17 meters at a distance of 10 km (Abe et al. 2006). While the NIRS wavelength range spans 0.76-2.25 μm, in reality the usable range is 0.85-2.1 μm due to fading detector sensitivity at the lower and higher wavelength ends of the spectrum. During operation NIRS made one-dimensional latitudinal scans with the target point fixed on Itokawa (Kitazato et al. 2008). This data set was acquired at phase angles between 0° and 38° with a 6 to 90 m² footprint (Abe et al. 2006). The NIRS data covers most of the surface of Itokawa with a total of 83,328 reflectance spectra measured. Figure 1 display the NIRS coverage with each data point representing the center coordinates of each NIRS footprint. The NIRS data set with several overlapping footprints can efficiently be used to create accurate global compositional maps of Itokawa. While NIRS limited wavelength range (0.85-2.1 μm) is not adequate for mineralogical analysis using existing spectral calibrations (e.g., Dunn et al. 2010) it complements that of the AMICA (0.38-1.0 μm) camera filter data. Results from AMICA observations show rough and smooth terrains of Itokawa containing angular rock fragments of several meters in size and buried pebble-sized particles respectively (Saito et al., 2006). AMICA acquired >1400 multispectral, high-resolution images (Ishiguro et al., 2010). Our goal is to combine datasets from the NIRS and AMICA instruments to create a global compositional map of Itokawa.

Note that Binzel et al. proposed that the surface of Itokawa was space weathered based on the interpretation of the spectra. It was not until the samples were returned and analyzed that this was found to be true.

The analysis of the NIR spectrum of Itokawa measured using ground-based telescopes suggested that its surface is similar to an ordinary chondrite, LL-type (Binzel et al., 2001). Binzel et al. (2001) proposed that the asteroid's surface is exposed to space weathering (a process of optical alteration) because the reflectance spectrum has a steep spectral slope and less pronounced absorption bands. The analysis of NIRS observations of Hyabusa also showed variations in band depth and spectral slope within different terrains on Itokawa. Abe et al. (2006) suggested that the observed heterogeneity might be present due to space weathering and particle size effects. However, the exact cause of the observed heterogeneity remains unknown as spectral data used for the space weathering interpretation was not photometrically corrected (Hiroi et al. 2008). Other observations from Hayabusa suggested the presence of albedo variations on Itokawa (Sasaki et al. 2007, Kitazato et al. 2008). However, ground-based observations revealed no rotational spectral variation. These Hayabusa results contradict the general assumption that most asteroids in the main belt, as well as the NEA population, have homogenous surface albedo/composition with the only exception so far being (4) Vesta (Gaffey, 1997; Reddy et al. 2012).

Here we present the first attempt to derive a set of calibration equations for extracting olivine and pyroxene composition from NIRS data, and to map their distribution on the surface of Itokawa. Our approach requires a direct correlation between diagnostic spectral band parameters (e.g., Band I and II centers and band area ratio or BAR) and laboratory XRD measurements of fayalite (olivine) and ferrosilite (pyroxene) mineral composition (mol.%) of ordinary chondrites. Reliable measurement of Band II spectral parameters from NIRS spectra are not possible due to the instrument's limited wavelength range (0.85-2.1 µm). We have therefore used only Band I parameters to derive the new calibration equations. We then tested the validity of these equations using laboratory spectra of meteorites with known composition and a small subset of uncalibrated NIRS spectral data of Itokawa.

## 2. Methodology

We selected 38 laboratory-measured reflectance spectra of H, L and LL chondrites available from the Reflectance Experiment Laboratory (RELAB). The same data set has been used by Dunn et al. 2010 to extract olivine and pyroxene compositions from NIR reflectance spectra (0.7-2.5 µm) of S-type asteroids. Table 1 presents a list of RELAB samples used in this study, representing the complete petrologic range of the equilibrated ordinary chondrite group (H, L, and LL) (Van Schmus and Wood, 1967). A detail description of the chondrite samples is described by Dunn et al. 2010.

The RELAB spectra (0.32-2.5 µm) are interpolated to the NIRS wavelength range (0.73-2.24 µm) and normalized to unity at 1.5 µm. Figure 2 shows a RELAB spectrum corresponding to the same spectrum resampled to NIRS wavelengths. The normalized spectra are divided by its convex hull in order to remove continuum (Fu et al., 2007). Since the automatic selection of the continuum is sensitive to the noise present in the remotely measured reflectance spectra, we have employed a Savitzky–Golay filter (Savitzky and Golay, 1964), as proposed by Lillesand et al. (2008), as a pre-processing step in our band parameters extraction routine. The Band I parameters are extracted from normalized and continuum-removed spectra. In this analysis only Band I center is used to derive calibration equations. Band I center is determined by fitting a third order polynomial to the bottom of the continuum-removed feature, and the minimum point of the polynomial is used as the band center.

NIRS includes the 2-µm absorption band up to 2.24 µm; however, we observed greater scatter beyond 2.1 µm due to a decrease in the detector sensitivity. As noted earlier, Band II spectral parameters cannot be computed from the NIRS data due to its limited wavelength range (Fig. 2, bottom panel). Thus our calibration method is based only on the extraction of Band I parameters. However, Band I parameters can be measured precisely for the NIRS spectra because the wavelength coverage in the Band I spectral range is complete compared to that of Band II. Band parameter values do show dependency on the selection of the continuum endpoints.

Dunn et al. (2010) presented a calibration method to extract olivine and pyroxene mineral composition by using spectral band parameters. They demonstrated that the Band I center is correlated with the abundance of FeO in olivine and pyroxene and successfully tested this relationship for 48 ordinary chondrite samples for which modal abundances were analysed using XRD. Our method of assessing FeO wt.% in Itokawa surface materials is based on the calibration algorithm originally developed by Dunn et al. (2010) for S-type asteroids. We selected 38 chondrite samples for our study from Dunn et al. (2010) since they were the best-characterized samples publicly available with XRD-measured abundance of fayalite in olivine (Fa) and ferrosilite in low-Ca pyroxene (Fs). Table 1 presents Fa and Fs values for the 38 ordinary chondrite samples used to derive calibration equations for the NIRS dataset. We computed the Band I center for the samples listed in Table 1 after resampling the spectra to the NIRS wavelength range. We found slight a discrepancy in Band I center between our study and Dunn et al. 2010. The maximum difference we found in Band I center calculation is 0.028. We have resampled the laboratory spectral data used by Dunn et al. (2010) to the NIRS wavelength range. The observed discrepancy in Band I center values might be due to the different spectral resolutions of the laboratory spectral data and NIRS data. Dunn et al. (2010) used a straight line continuum by selecting the data points on either side of Band I whereas, we removed continuum by dividing the normalized spectra by its convex hull (Section 2). The different methods used for continuum removal could also be a reason of observed discrepancy in Band I values. However, the observed discrepancies in Band I center values do not affect mineralogical composition.

We computed the Band I center for the samples listed in Table 1 after resampling the spectra to the NIRS wavelength range. We plotted the measured Fa and Fs values as a function of Band I center as shown in Figs. 3 (a) and 4 (a). These figures demonstrate that a clear correlation exists between measured Fa and Fs values and Band I center. We fitted a second-order polynomial to these relationships in order to represent them in the form of equations which can be generalized to the NIRS dataset:

Fa = -790.2 x (Band I Center)$^2$ + 1649.7x (Band I Center) – 830.3         (Eq. 1)

Fs = -521.7 x (Band I Center)$^2$ + 1094.8x (Band I Center) – 549.1         (Eq. 2)

The fayalite and ferrosilite contents derived from Eq. (1) and (2) show a correlation between derived and measured values of 93% and for 92%. The standard deviation computed for the differences between the measured and estimated Fa and Fs values are 1.5 and 1.1 mol. %, respectively. These results for RELAB data interpolated to the NIRS wavelength range are shown in Figs. 3b and 4b. The maximum scatter observed from Figs 3b and 4b is between 5% and 6% for the meteorite samples listed in Table 1.

## 3. Validation of Equations for laboratory measured dataset

In an effort to validate the derived equations we selected spectra of a sample of nine meteorites (LL and L chondrites) for which Fa and Fs modal abundances are available but were not used to create the original calibration equations, and applied our calibration equations to these samples to derive Fa and Fs values. Table 2 presents the laboratory measured Fa and Fs values and spectrally-derived values using Eq's. 1 and 2 along with the difference between these values. The root mean square error of the spectrally derived band I center is 0.02. We found that the absolute difference between the laboratory measured and spectrally estimated values is <1.7 mol.% and the standard deviation computed for the difference between the measured and estimated Fa, Fs values is ~1 mol.% assuming the data from the Ausson L5 is omitted.

## 4. Validation of Equations for Hayabusa NIRS

The Hayabusa mission phase consisted of three subphases depending on the distance to the asteroid (Kitazato et al. 2008). These include the Gate-Position (GP) phase (20-8 km), Home-Position (HP) phase (~8 km), and the descent and touchdown (TD) phase during which the spacecraft attempted to collect surface samples. These data sets are archived on the PDS node and available online. We selected a data set recorded on 10/10/2005 from home position to test our new calibration equations. This data set consists of a total of 1589 spectra with most of the footprints covering the surface area between 45°N and 45°S. Note that the data are not photometrically calibrated and hence have weaker constraints on the spectral slope and band depth (Sanchez et al. 2012). Figure 5 shows a few sample spectra from the NIRS instrument. We extracted Band I center from the archived data after applying a Savitsky–Golay filter function, as explained in section 2. We found a maximum deviation of 0.02 μm in Band I center calculation by gradually changing frame size of the Savitsky-Golay filter function. We found that the maximum deviation observed in Band I center position actually does not change its mineralogy and is considered well within error bars. We applied Eq's. 1 and 2 to this data set and generated preliminary Fa and Fs maps of the Itokawa surface (Fig. 6). Photometric corrections have not been applied to this data set, and absolute values may differ after applying these corrections because the observing geometry affects the position of the absorption bands, band depth, and spectral slope. The data displayed in Figs. 6 (a) and (b) are color-coded based on their Fa and Fs values. Using NIRS data acquired on 10/10/2005 we found Fa values ranging from ~25 to ~30 and Fs values ranging from ~21 to ~25, which is consistent with an LL-type composition for Itokawa.

## 5. Summary and future work

We derived equations to constrain the olivine and pyroxene composition of asteroid Itokawa using Hayabusa NIRS spectrometer data. Using 38 laboratory-measured reflectance spectra of H, L and LL chondrites, we established a relationship between Band I center and their mafic silicate compositions in a similar way as Dunn et al. 2010. We found a correlation cofficient >90% between mafic silicate compositions estimated by our method and XRD-measured values. We validated the derived calibration equations by applying them to nine laboratory measured spectra of L and LL type ordinary chondrites with known composition. The spectrally-derived values for fayalite and forsterite are consistent with those measured in the lab. As a follow-up test, we applied the derived calibration equations to a subset of NIRS spectra archived on the PDS. The fayalite and ferrosilite values we estimate using the derived calibration equations confirms the LL type ordinary chondrite composition of Itokawa .

As a further refinement step, a photometric model will be developed to correct the NIRS data archived on the PDS Small Bodies node for the effect of illumination conditions. Photometric modeling is important prior to extraction of spectral band parameters from any spacecraft data. Itokawa has a very irregular shape and the NIRS spectral data were recorded under widely varying photometry conditions. Hence, the NIRS spectral data must be photometry corrected in order to produce an appropriate interpretation of spectral variations.

Our future work will include construction of global Fa and Fs maps using the methods developed in this study. These maps will be valuable for identifying subtle compositional variations on Itokawa. We will also consider modifying the algorithm developed in this study for use in analyzing other ground based and spacecraft mission data sets.

**Acknowledgments**

This research work was supported by NASA Planetary Mission Data Analysis Program Grant NNX13AP27G. The authors would like to thank the two referees, Paul Abell and the anonymous reviewer, for their detailed comments and positive suggestions. The authors would like to thank Bruce Gary (Hereford Arizona Observatory) for his helpful comments to improve the manuscript.

**Figure Captions**

**Figure 1:** Global distribution of the NIRS observations projected onto a cylindrical equi-distant map. Each data point corresponds to the center coordinates of a NIRS footprint. The location of each footprint center is provided in the label files of the NIRS data in the PDS. The AMICA image mosaic of Itokawa used as a background map is from Stooke et al. (2012).

**Figure 2:** The RELAB and NIRS reflectance spectra with continuum line fitted for extracting band parameters. The continuum line consists of straight line segments fitted to the reflectance spectra normalized at 1.5 μm. The band center position is marked with a star. (a) Tuxtuac meteorite (Table 1, RELAB id: MT-HYM-080) spectrum plotted corresponding to full wavelength coverage. The Band I and Band II parameters can be extracted from such spectra due to the full wavelength coverage; (b) the reflectance spectrum of Tuxtuac meteorite is interpolated to the NIRS wavelength range. Band II parameters cannot be extracted confidently from such spectra due to limited wavelength coverage; (c) remotely measured NIRS reflectance spectrum of Itokawa measured on 10/10/2005 from home position (HP) phase. Band II parameters cannot be extracted confidently from the NIRS data due to its limited wavelength range and low detector sensitivity beyond 2.1 μm.

**Figure 3:** (a) Laboratory-measured mol. % of fayalite (Fa) in olivine plotted as a function of Band I center for 38 chondrite samples listed in Table 1. A second-order polynomial is fitted (solid black line). (b) laboratory-measured versus spectrally-derived Fa values. The correlation coefficient is 0.937.

**Figure 4:** (a) Laboratory-measured mol% of ferrosilite (Fs) plotted as a function of Band I center for 38 chondrite samples listed in Table 1. A second-order polynomial is fitted (solid black line). (b) laboratory-measured versus spectrally-derived Fs values. The correlation coefficient is 0.929.

**Figure 5:** The NIRS reflectance spectra measured on 10/10/2005 from home position (HP) phase. We show five consecutive reflectance spectra with center coordinates from 14.5°E to 25.1°E and 18.1°S to 23.9°S. The derived Fa and Fs values (Eq's. 1 and Eq. 2) are in the range 27.9 to 29.2 and 22.9 to 23.9, respectively.

**Figure 6:** A total of 1583 reflectance spectra of Itokawa measured on 10/10/2005 from home position (HP) phase. Fayalite (Fa) and ferrosilite (Fs) values are plotted corresponding to the corner coordinates provided in the label file of each NIRS data file in the PDS; (a) the fayalite distribution is derived using Eq.1. Fa values are ranging from ~25 to ~30; (b) the ferrosilite distribution is derived using Eq. 2. Fs values are ranging from ~21 to ~25. These spectra are not corrected for photometry and are only shown as a test case. Therefore, variations in Fs and Fa values are dominated by changing viewing conditions. Six NIRS data files were omitted from this sequence (folder 20051010 in the PDS) because of abnormal spectral shape or very low reflectance value

(maximum reflectance measured at ~0.01 in the whole spectrum). The AMICA image mosaic of Itokawa used as a background map is from Stooke et al. (2012).

Table 1: List of the meteorite samples used in this study.

| Meteorite | RELAB ID | Type | Fa | Fs |
|---|---|---|---|---|
| Benares (a) | MT-HYM-083 | LL4 | 23.0 | 28.7 |
| Hamlet | MT-HYM-075 | LL4 | 22.0 | 26.4 |
| Aldsworth | MT-HYM-077 | LL5 | 23.6 | 28.2 |
| Olivenza | MT-HYM-085 | LL5 | 24.2 | 29.9 |
| Paragould | MT-HYM-079 | LL5 | 23.0 | 27.6 |
| Tuxtuac | MT-HYM-080 | LL5 | 25.3 | 30.4 |
| Bandong | TB-TJM-067 | LL6 | 24.6 | 30.4 |
| Cherokee Springs | TB-TJM-075 | LL6 | 23.0 | 28.2 |
| Karatu | TB-TJM-077 | LL6 | 25.4 | 30.9 |
| Saint-Severin | TB-TJM-145 | LL6 | 23.9 | 29.6 |
| Attara | TB-TJM-065 | L4 | 19.5 | 23.1 |
| Bald Mountain | TB-TJM-102 | L4 | 19.2 | 22.8 |
| Rio Negro | TB-TJM-081 | L4 | 19.9 | 24.4 |
| Rupota | TB-TJM-121 | L4 | 20.0 | 24.4 |
| Ausson | MT-HYM-084 | L5 | 20.5 | 23.9 |
| Guibga | TB-TJM-134 | L5 | 20.4 | 24.6 |
| Malakal | TB-TJM-109 | L5 | 20.8 | 24.7 |
| Messina | TB-TJM-099 | L5 | 20.6 | 24.4 |
| Apt | TB-TJM-064 | L6 | 21.5 | 24.9 |
| Aumale | TB-TJM-101 | L6 | 20.3 | 24.3 |
| Karkh | TB-TJM-137 | L6 | 21.1 | 24.7 |
| Kunashak | TB-TJM-139 | L6 | 20.4 | 24.5 |
| Kyushu | TB-TJM-140 | L6 | 20.4 | 24.3 |
| New Concord | TB-TJM-130 | L6 | 20.3 | 24.2 |
| Farmville | TB-TJM-128 | H4 | 15.7 | 17.5 |
| Kabo | TB-TJM-136 | H4 | 16.2 | 18.1 |
| São Jose do Rio Preto | TB-TJM-082 | H4 | 16.5 | 18.8 |
| Allegan | TB-TJM-104 | H5 | 15.6 | 17.6 |
| Ehole | TB-TJM-074 | H5 | 17.0 | 19.3 |
| Itapicura Mirim | TB-TJM-097 | H5 | 16.3 | 18.2 |
| Lost City | TB-TJM-129 | H5 | 16.1 | 18.6 |
| Pribram | TB-TJM-143 | H5 | 16.6 | 18.4 |
| Schnectady | TB-TJM-083 | H5 | 16.6 | 18.9 |
| Andura | TB-TJM-088 | H6 | 16.5 | 19.2 |
| Butsura | TB-TJM-069 | H6 | 16.8 | 19.0 |
| Canon City | TB-TJM-131 | H6 | 16.6 | 19.1 |
| Guareña | TB-TJM-094 | H6 | 16.8 | 19.1 |
| Ipiranga | TB-TJM-135 | H6 | 16.5 | 18.6 |

**Table 2: Test data from RELAB with their measured and spetrally-derived Fa, Fs values.**

| Meteorite | Type | RELAB ID | Band I Center | Fa Measured | Fa derived | Fa Difference | Fs Measured | Fs Derived | Fs Difference |
|---|---|---|---|---|---|---|---|---|---|
| Attarra | L4 | TB-TJM-065 | 0.95 | 23.1 | 23.9 | -0.8 | 19.5 | 20.1 | -0.6 |
| Hamlet | LL4 | MT-HYM-075 | 0.97 | 26.4 | 26.9 | -0.5 | 22.0 | 22.2 | -0.2 |
| Bernares | LL4 | MT-HYM-083 | 1.02 | 28.7 | 30.2 | -1.5 | 23.0 | 24.7 | -1.7 |
| Ausson | L5 | MT-HYM-084 | 0.92 | 23.9 | 20.1 | 3.8 | 20.5 | 17.5 | 3.0 |
| Aldsworth | LL5 | MT-HYM-077 | 0.97 | 28.2 | 26.9 | 1.3 | 23.6 | 22.2 | 1.4 |
| Paragould | LL5 | MT-HYM-079 | 0.97 | 27.6 | 26.9 | 0.7 | 23.0 | 22.2 | 0.8 |
| Tuxtuac | LL5 | MT-HYM-080 | 1.04 | 30.4 | 30.5 | -0.1 | 25.3 | 25.0 | 0.3 |
| Olivenza | LL5 | MT-HYM-085 | 1.02 | 29.9 | 30.2 | -0.3 | 24.2 | 24.7 | -0.5 |
| Apt | L6 | TB-TJM-064 | 0.95 | 24.9 | 23.9 | 1.0 | 21.5 | 20.1 | 1.4 |

Figure 1.

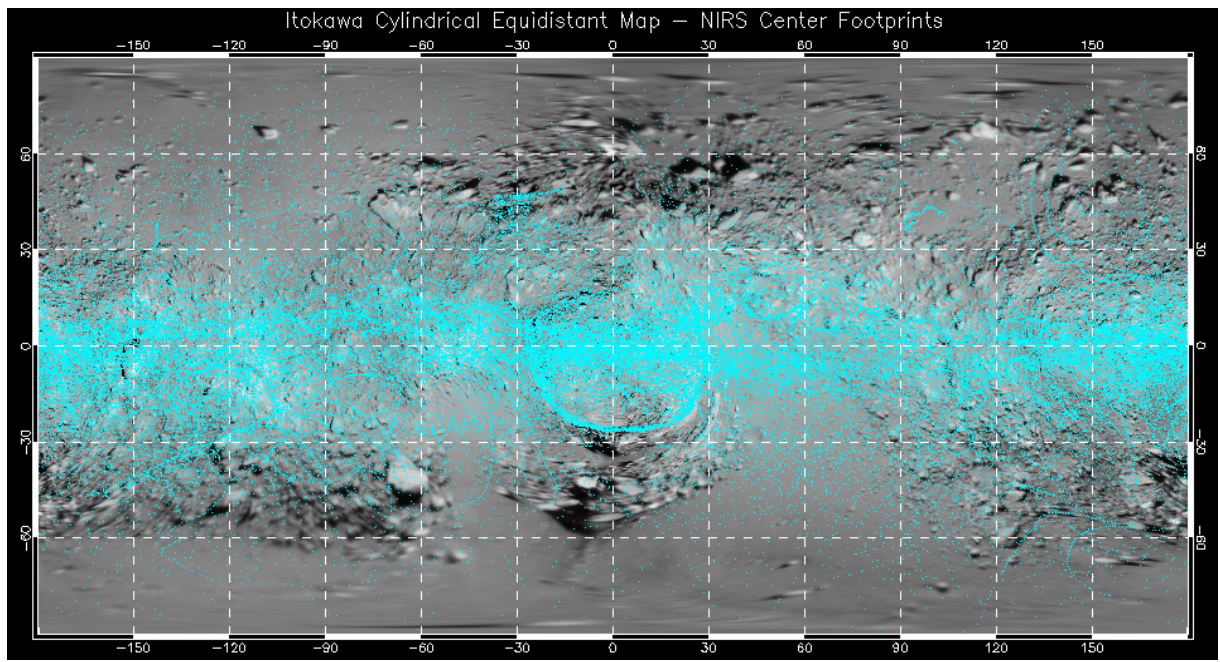

Figure 2.

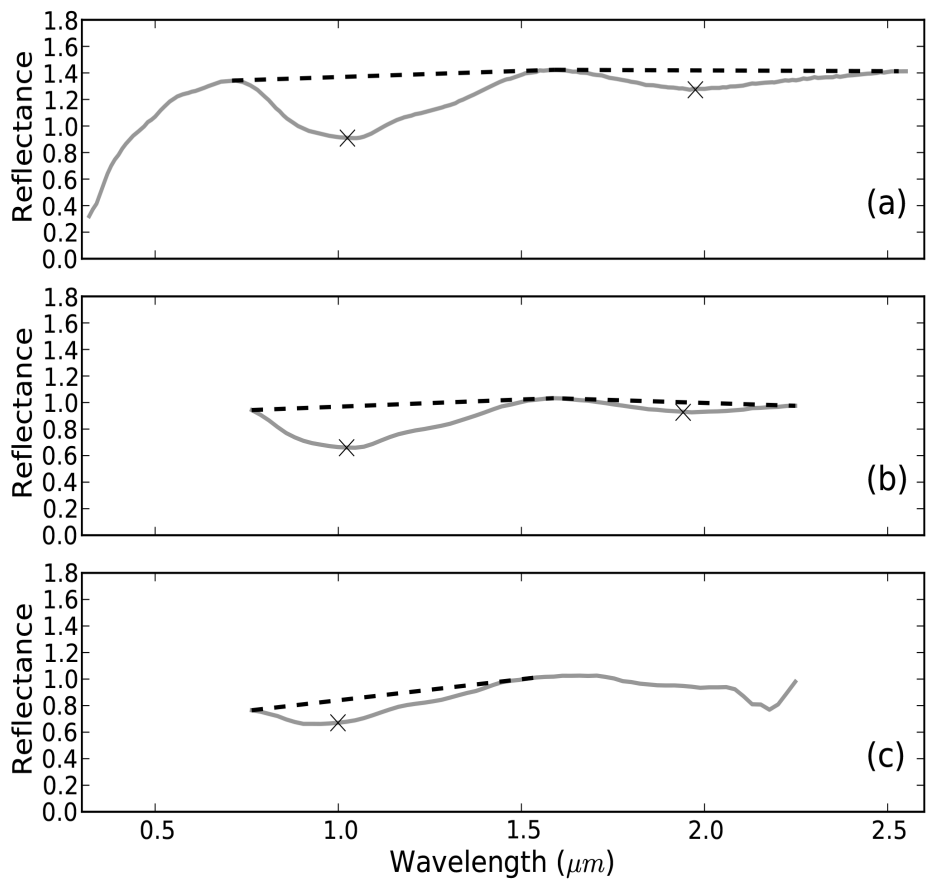

Figure 3.

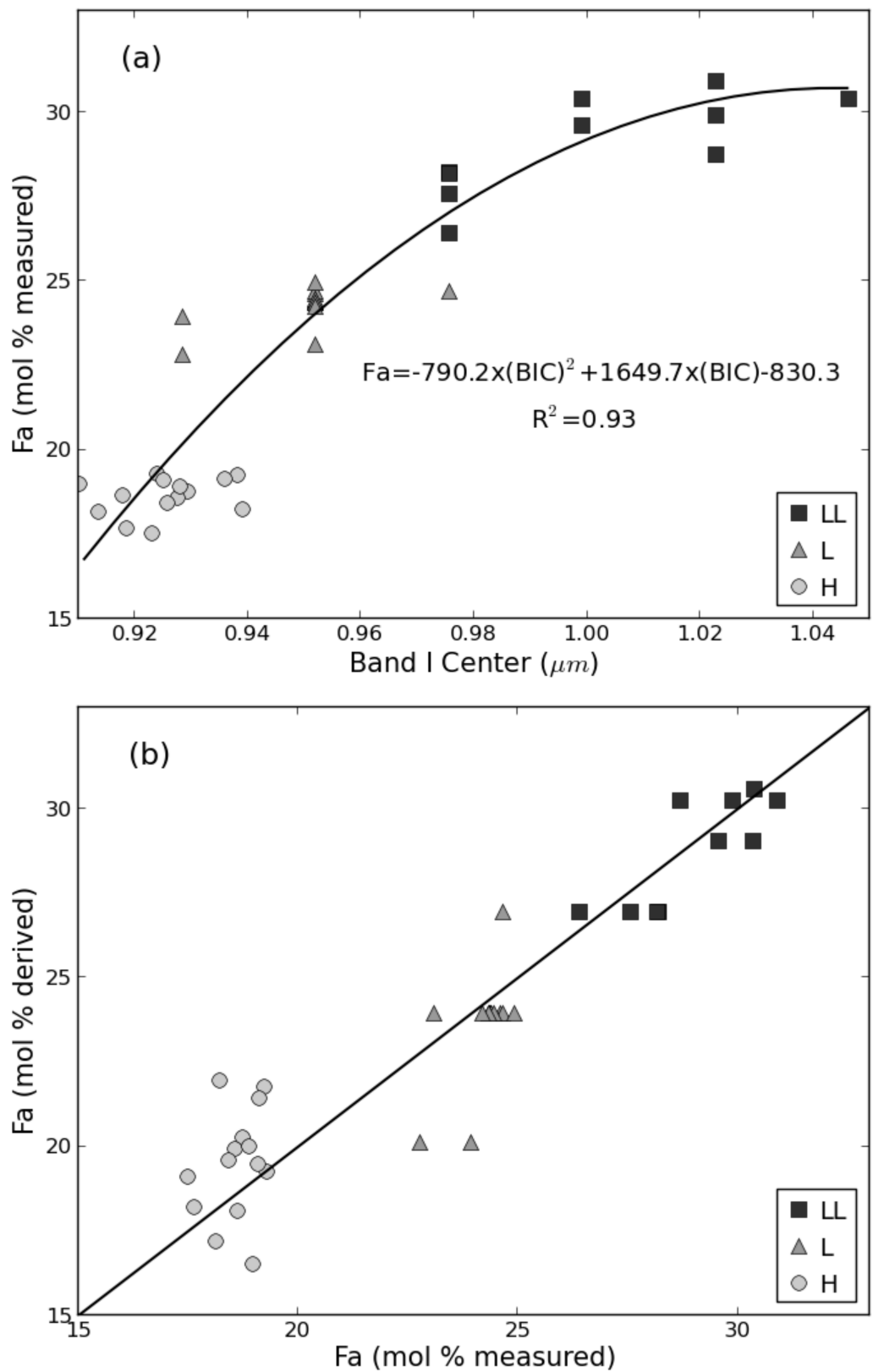

Figure 4.

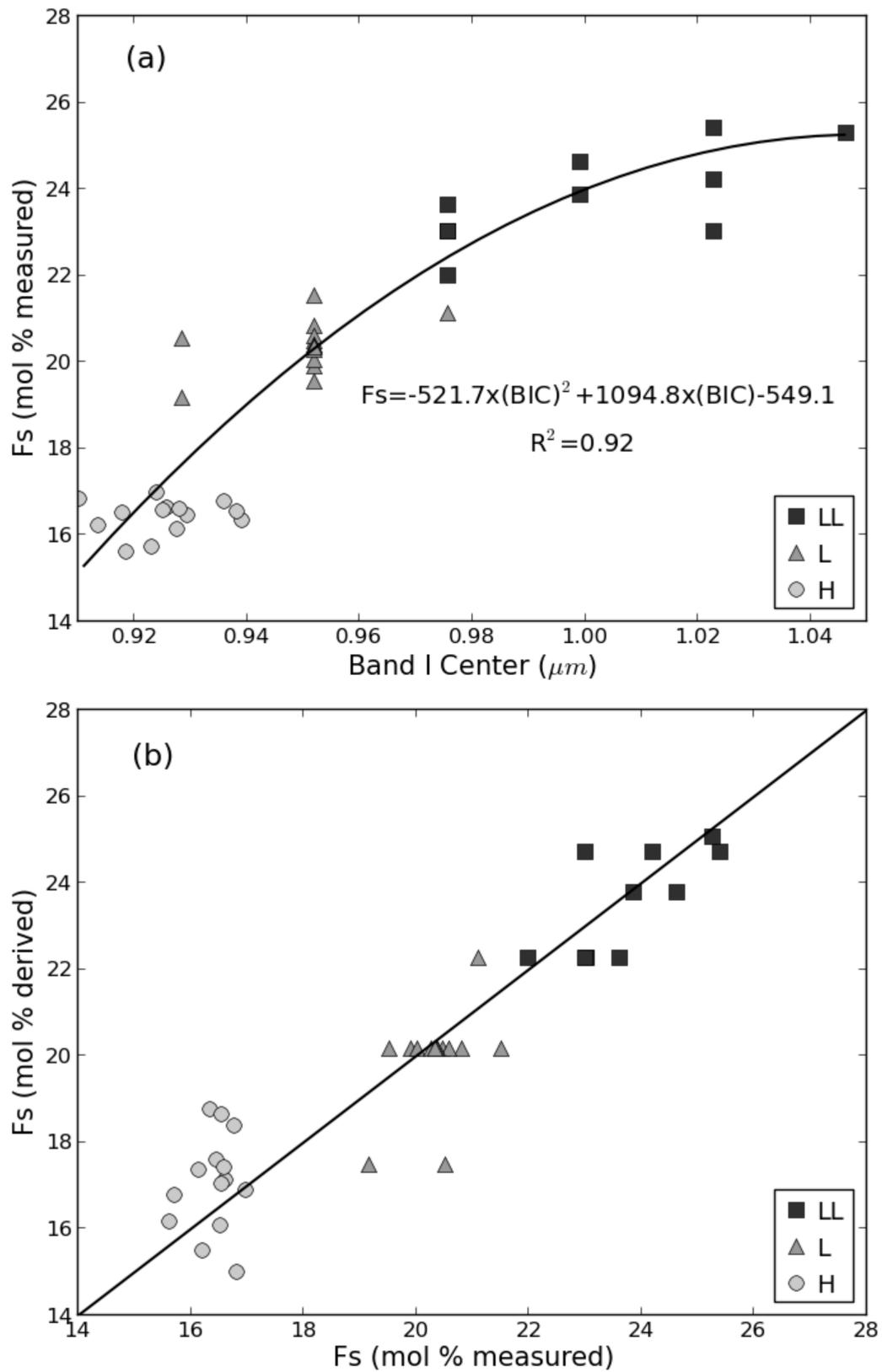

Figure 5.

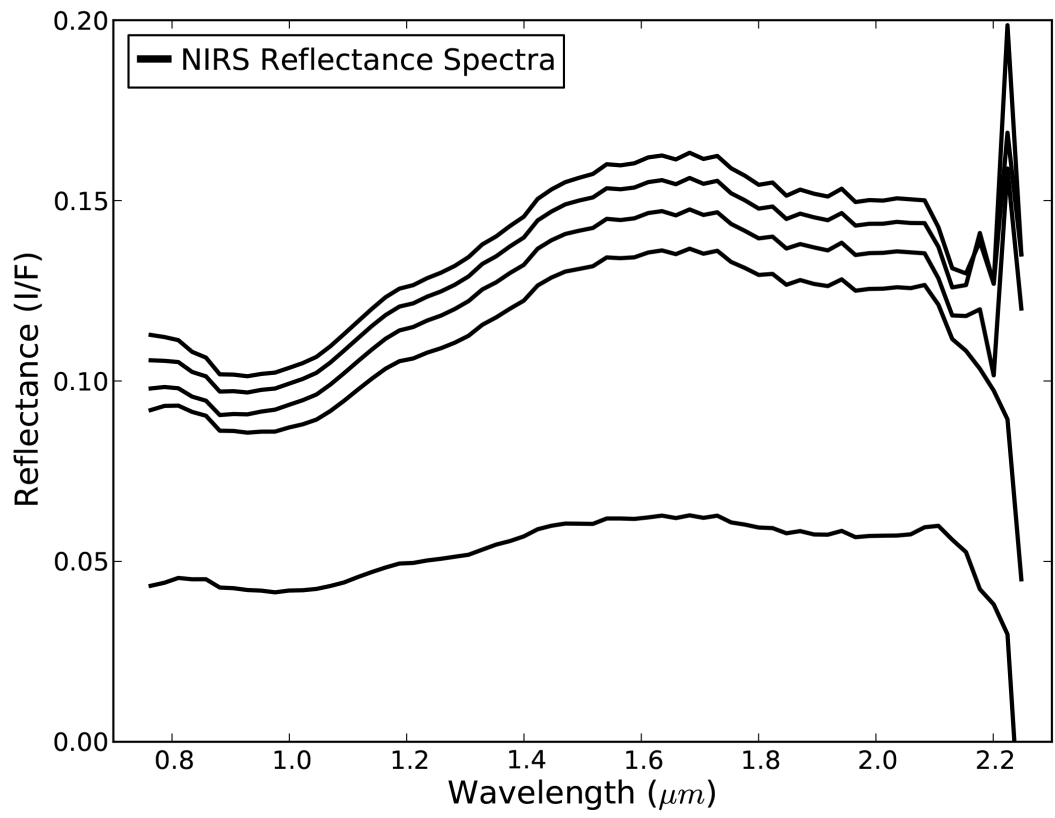

Figure 6.

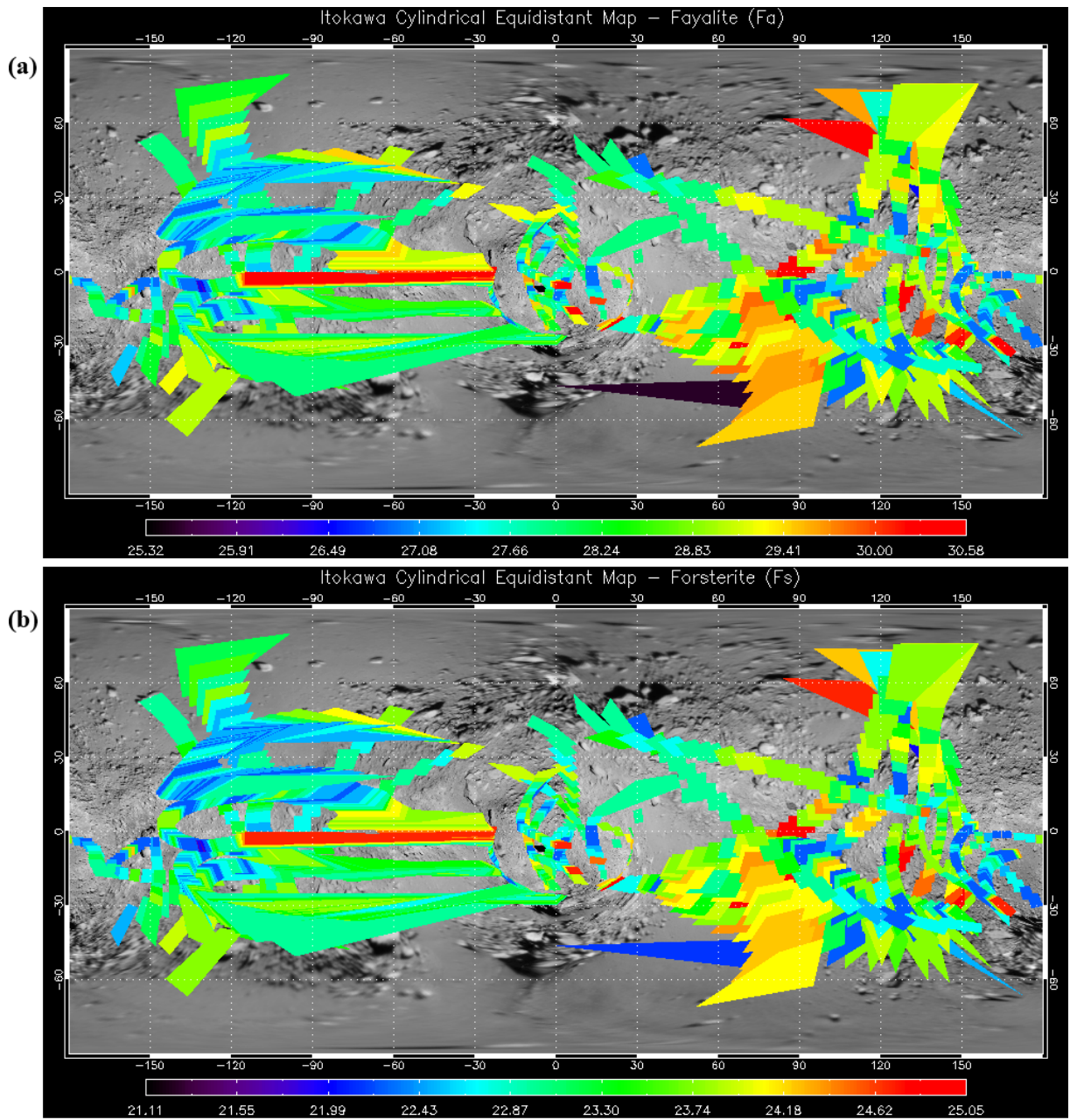